\documentclass[prl, twocolumn]{revtex4}

\usepackage{amsmath}
\usepackage{amssymb}
\usepackage{graphicx}
\usepackage{subfigure}
\usepackage{color}
\usepackage{psfrag}

\newcommand{\erdosrenyi}{Erd\H{o}s--R\'{e}nyi}
\newcommand{\ave}[1]{\left \langle #1 \right \rangle}

\newcommand{\I}{\mathcal{I}}
\newcommand{\Sus}{\mathcal{S}}
\newcommand{\PE}{\mathcal{P}}
\newcommand{\A}{\mathcal{A}}
\newcommand{\order}{\mathcal{O}}
\newcommand{\littleo}{o}

\newcommand{\binomial}[3]{\mathrm{Bi}(#1,#2,#3)}

\begin{document}

\psfrag{At}[][][2.5][0]{$\A$}
\psfrag{Pr}[][][2.5][0]{$\PE$}
\psfrag{T*}[][][2.5][0]{$T^*$}

\title{Predicting the size and probability of epidemics in a
  population with heterogeneous infectiousness and susceptibility}

\author{Joel C. Miller} \affiliation{Mathematical Modeling \& Analysis
  Group and Center for Nonlinear Studies, MS B284, Los Alamos National
  Laboratory} \email{jomiller@lanl.gov} \thanks{Los Alamos Report
  LA-UR-06-8193} \date{\today}

\begin{abstract}
  We analytically address disease outbreaks in large, random networks
  with heterogeneous infectivity and susceptibility.  The
  transmissibility $T_{uv}$ (the probability that infection of $u$
  causes infection of $v$) depends on the infectivity of $u$ and the
  susceptibility of $v$.  Initially a single node is infected,
  following which a large-scale epidemic may or may not occur.  We use
  a generating function approach to study how heterogeneity affects
  the probability that an epidemic occurs and, if one occurs, its
  attack rate (the fraction infected).  For fixed average
  transmissibility, we find upper and lower bounds on these. An
  epidemic is most likely if infectivity is homogeneous and least
  likely if the variance of infectivity is maximized.  Similarly, the
  attack rate is largest if susceptibility is homogeneous and smallest
  if the variance is maximized.  We further show that heterogeneity in
  infectious period is important, contrary to assumptions of previous
  studies.  We confirm our theoretical predictions by simulation.  Our
  results have implications for control strategy design and
  identification of populations at higher risk from an epidemic.
\end{abstract}

\maketitle

The spread of infectious disease is a problem of great
interest~\cite{andersonMay}.  Much work has focused on how diseases
spread in networks of human, animal, or computer
interactions~\cite{meyers:directed,meyers:contact,newman:spread,kao,madar,serrano:prl,boguna:scalefree,pastor-satorras:scale-free}.
The transmissibility, the probability that an edge transmits
infection, has a network-dependent threshold
(which can be zero) corresponding to a second order phase transistion
above which an epidemic may happen and below which epidemics are not
found.  Ideally an intervention reduces the transmissibility or
modifies the network to raise
the threshold so that epidemics cannot occur.  Most study has focused
on determining the threshold value under varying
assumptions~\cite{madar,serrano:prl,boguna:scalefree,pastor-satorras:scale-free,dodds:contagionprl,dodds:contagionJTB}
in order to design an optimal intervention.

For many diseases and networks, it is impractical to reduce
transmissibility sufficiently to eliminate the possibility of an
epidemic.  An intervention strategy should therefore optimize
competing goals: minimize social cost, reduce the probability a
large-scale epidemic occurs, and reduce the attack rate (fraction
infected) if an epidemic does occur.  Recently the probability and
attack rate have been
investigated~\cite{meyers:directed,meyers:contact,meyers:sars,newman:spread,newman:threshold},
but none of these has systematically investigated the effect of
heterogeneity in transmissibilities.  Heterogeneities can result from
variations in the application of interventions or from natural
differences in the population such as variation in recovery time.  It
is often assumed that this special case can be mapped without loss of
generality to recovery of all individuals after a single time
step~\cite{newman:spread,newman:structurereview,madar,serrano:prl,boguna:scalefree,pastor-satorras:scale-free}
and so the number of new cases from a single case is distributed
binomially.  However, it may be inferred from~\cite{hastings:series}
that this assumption is false.  We have recently become aware of
independent work~\cite{kenah} which shows that recovery time
heterogeneity reduces the epidemic probability, but has no effect on
the attack rate.  In this Letter, we consider how generic
heterogeneities affect the epidemic probability and the attack rate if
an epidemic occurs.


The epidemics we study spread on random networks of $N$ nodes with
degree distribution given by $P(k)$ where $k$ is the degree.
We use the SIR model~\cite{andersonMay}: nodes are divided into
susceptible, infectious, and recovered classes.  We modify the model
to include heterogeneities.  An infectious node $u$ with infectivity
$\I_u$ connected to a susceptible node $v$ with susceptibility
$\Sus_v$ infects $v$ with probability equal to the transmissibility
$T_{uv}(\I_u,\Sus_v)$ of the edge.  Infectious nodes recover and are
no longer susceptible.  The outbreak begins with a single infection
(the \emph{index case}) which spreads to neighboring nodes.  If an
epidemic occurs, the eventual number infected is $\order(N)$,
otherwise the outbreak is localized.  $\I$ and $\Sus$ can be quite
arbitrary, \emph{e.g.}, $\I$ may be a vector representing time of
infection, level of virus shedding, frequency of handwashing, etc.
The form of $T_{uv}$ can also be quite general: it need only be
integrable and bounded in $[0,1]$.



\begin{figure}
\input{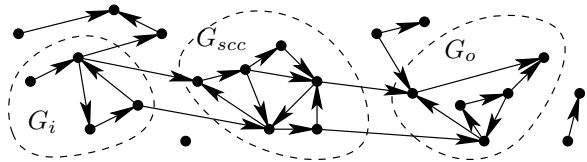}
\caption{Schematic representation of $G_i$, $G_{scc}$, and $G_o$.  All
  nodes in $G_{scc}$ can reach any other node in $G_{scc}$.  }
\label{fig:Gscc}
\end{figure}

The spread of an epidemic on a network with heterogeneous infectivity
and susceptibility is equivalent to a special case of directed
percolation for which the probability of retaining an edge depends on
both the base and target node.  In this formalism, infection spreads
to the out-component of the index
case~\cite{meyers:directed,meyers:contact}.  If the disease has
sufficiently high transmissibility a single giant strongly connected
component $G_{scc}$ exists~\cite{broder}, occupying a fixed fraction
of the network as $N \to \infty$.  The set of nodes not in $G_{scc}$,
but from which $G_{scc}$ can be reached is denoted $G_i$, while the
set of nodes not in $G_{scc}$ but reachable from $G_{scc}$ is denoted
$G_o$ [$i$ for `in' and $o$ for `out'] as demonstrated in
figure~\ref{fig:Gscc}.  If the index case is in $G_i\cup G_{scc}$ an
epidemic occurs, infecting all of $G_o \cup G_{scc}$ and very few
other nodes.  In the limit $N\to\infty$ the probability of an epidemic
is the probability the index case is in $G_i \cup G_{scc}$ and the
attack rate is the fraction of nodes in $G_o \cup
G_{scc}$~\footnote{This statement is true for the probability without
  requring $N\to\infty$, but if the initial infection is in $G_i$, a
  few nodes outside of $G_o\cup G_{scc}$ may also be infected, leading
  to a $\littleo(1)$ correction to the attack rate.}.  We use $\PE$ and $\A$ to denote the limiting epidemic
probability and attack rate.  In general the sizes of $G_i$ and $G_o$
may differ significantly so $\PE \neq \A$.  This contrasts with the
case of homogeneous transmissibility where the problem can be mapped
to undirected bond
percolation~\cite{grassberger:percolation,cardy:percolation,newman:spread}
and $\PE=\A$.

We develop a general theory to find $\PE$ allowing both infectivity
and susceptibility to be heterogeneous.  Generating function
approaches~\cite{gf} have been used to study disease spread both inside the
body~\cite{wahl:burst} or in
society~\cite{newman:spread,meyers:contact}.  We modify
these approaches to calculate $\PE$ based on the distribution of $\I$
and $\Sus$.  Holding the average transmissibility fixed, we then use
Jensen's inequality to find distributions which give upper and lower
bounds on $\PE$.  Because $G_o$ and $G_i$ interchange roles if edge
directions are reversed, $\A$ is calculated in the same manner.  Our
predictions are confirmed through simulations on a large \erdosrenyi{}
network.

Each node $u$ has an infectivity $\I_u$ and a susceptibility $\Sus_u$
chosen from independent distributions given by $P(\I_u)$ and
$P(\Sus_u)$.  Given the infectivity $\I_u$ of $u$, the relation
$T_{uv}(\I_u,\Sus_v)$, and the distribution $P(\Sus)$, we define the
out-transmissibility of $u$ as
\begin{equation}
T_o(u) = \int T_{uv}(\I_u,\Sus_v) P(\Sus_v) \, d\Sus_v \, .
\label{eqn:To}
\end{equation}
From~\eqref{eqn:To} and the distribution $P(\I)$, we know
the distribution $P_o(T_o)$.  We similarly define the
in-transmissibility $T_i$ and its distribution $P_i(T_i)$.
$P_o$ and $P_i$ must yield the same average, but not all pairs $P_o$
and $P_i$ with the same average are consistent.  For each $P_o$ there
exists at least one $P_i$ and \emph{vice versa}.  Henceforth we
consider just $P_i$ and $P_o$, and do not use $P(\Sus)$ and $P(\I)$.

We choose the index case $u_0$ uniformly from the population.  We
classify infected cases by their generation, measuring the number of
infectious contacts in the chain between them and $u_0$ (generation
$0$).  We note that the generation time need not be fixed: generations
may overlap in time, changing the temporal dynamics but not affecting
our results.

Our class of random networks is defined by the Molloy--Reed
algorithm~\cite{molloyReed}.  Short cycles are rare. The neighborhood
of $u_0$ is tree-like on successively longer length-scales as $N \to
\infty$.  Consequently, $\PE$ equals the probability that the
transmission chains in an infinite tree are infinite.

We define a probability generating function $f(x)$ for the number of
infected nodes in generation~$1$:
\[
f(x) = p_0 + p_1 x + \cdots + p_j x^j + \cdots \, ,
\]
where $p_j$ is the probability that the index case directly infects
$j$ neighbors.  The index case has degree $k$ with probability $P(k)$ and
thus $p_j$ is given by
\[
p_j = \sum_{k=j}^\infty P(k) \int_0^1 \binomial{k}{j}{T_o}
P_o(T_o) \,
dT_o \, ,
\]
where $\binomial{k}{j}{T_o}$ is the likelihood of $j$
successful trials from $k$ attempts, each with probability $T_o$.
Note that $p_j$ depends on the distribution $P_o$ but not $P_i$.

In subsequent generations, the probability that a node is infected is
proportional to its degree.  Early in the epidemic an infected node
with degree $k$ has $k-1$ susceptible neighbors because the source of
its infection cannot be reinfected.  As such the probability $q_j$
that this individual infects $j$ neighbors is 
\[
q_j = \frac{1}{\ave{k}} \sum_{k=j+1}^\infty k P(k) \int_0^1
\binomial{k-1}{j}{T_o}
 P_o(T_o)\, dT_o \, .
\]
where $\ave{\cdot}$ denotes the expected value.  We let $h(x)=\sum
q_j x^j$ be the generating function for the number of new cases
caused by a non-index case.  The generating function for the number of
infections caused by $n$ non-index cases is $[h(x)]^n$.  Consequently it
may be shown that the generating function for the number of infections
in generation $g>0$ is given by
\[
f(h^{g-1}(x)) \, ,
\]
where $h^{g-1}$ denotes composition of $h$ with itself $g-1$ times.
For later use we rearrange $f$ and $h$ as
\begin{align}
f(x) &= \int_0^1 P_o(T_o)\sum_{k=0}^\infty [1+T_o(x-1)]^k P(k)  \,
dT_o \, , \label{eqn:fform}\\
h(x) &= \int_0^1 \frac{P_o(T_o)}{\ave{k}}\sum_{k=1}^\infty [1+T_o(x-1)]^{k-1} k P(k)
   \, dT_o \,.  \! \label{eqn:hform}
\end{align}

The extinction probability is $\lim_{g\to \infty} f(h^{g-1}(0))$.  To
calculate this, we find $\lim_{g\to\infty} h^{g-1}(0)$ which is a
solution to $x=h(x)$.  At most
two solutions exist in the interval $[0,1]$, one of which is $x=1$.
If no other solution exists then $x=1$ is a stable fixed point and
$\PE=0$.  Otherwise the iteration converges to $x_0<1$ and
\[
\PE = 1-f(x_0) \,.
\]
Because $f$ and $h$ are independent of $P_i$, $\PE$ is unaffected by
heterogeneities in susceptibility.

We now seek distributions $P_o$ maximizing or minimizing $\PE$ subject
to $\ave{T}=T^*$.  In their investigation of recovery time
heterogeneities, \cite{kenah} showed that the probability is maximized
if recovery times are identical.  We generalize
this to arbitrary sources of heterogeneities in infectivity, using a
similar proof.  For notational
convenience we use $\delta^*(T)$ to denote the $\delta$-function
$\delta(T-T^*)$, set
\[
\hat{h}(T,x) = \frac{1}{\ave{k}}\sum_{k=1}^\infty [1+T(x-1)]^{k-1}
kP(k) \, ,
\]
and rewrite~\eqref{eqn:hform} to explicitly show that $h$ depends on
$P_o$
\[
h[P_o](x) = \int_0^1 \hat{h}(T_o,x)P_o(T_o)\, dT_o \, .
\] 
We similarly define $f[P_o](x)$.  
Because $\hat{h}$ is a convex function of $T$, Jensen's inequality
shows $P_o=\delta^*$ minimizes $h[P_o](x)$.  We denote the smallest
root of $x=h[\delta^*](x)$ by $x_1$.  For $x<x_1$ and any $P_o$, we
have $x<h[\delta^*](x) \leq h[P_o](x)$.  Thus the root $x_0$ of
$x=h[P_o](x)$ satisfies $x_1 \leq x_0$, so $x_0$ is minimized if
$P_o=\delta^*$.

Similar calculations show $f\left[\delta^*\right](x) \leq
f[P_o](x)$ for all $P_o$.  Further, $f\left[\delta^*\right](x)$ is
an increasing function of $x$.  Thus the extinction probability
$f[P_o](x_0)$ is minimized by $P_o=\delta^*$.  So homogeneous
infectivity maximizes $\PE$.

In addition, we find a new lower bound.  Jensen's inequality also
implies that fixing $\ave{T}=T^*$ but increasing $\ave{T^2}$ reduces
$\PE$.  Consequently, $\PE$ is minimized by $P_o(T_o) =
(1-T^*)\delta(T_o) + T^*\delta(T_o-1)$.

Thus we have shown that an epidemic is most likely if $T_o$ is
homogeneous and least likely if its variance is maximized.
Analogously the attack rate is largest if $T_i$ is homogeneous, and
smallest if its variance is maximized.

We expect that a threshold value of $\ave{T}$ exists above which
epidemics can occur [$x=h(x)$ has two roots] and below which they
cannot.  Allowing $\ave{T}$ to vary by continuously changing
$P_o$, the fixed point $x=1$ of $x=h(x)$ bifurcates into two when
$h'(1)=1$.  We find
\[
h'(1) 
      = \frac{\ave{T}\ave{k^2-k}}{\ave{k}} \, .
\]
So the epidemic threshold is $\ave{T} = \ave{k}/\ave{k^2-k}$,
generalizing the results of~\cite{newman:spread,meyers:contact}.

We confirm our predictions by comparison with simulations on
an \erdosrenyi{} network with $100000$ nodes and $\ave{k}=4$.  As $N
\to \infty$, \erdosrenyi{} networks with fixed average degree have Poisson
degree distribution and
\[
h(x) = f(x) = \int_0^1 \exp[\ave{k}T_o(x-1)] P_o(T_o) \, dT_o\, .
\]

\begin{figure}
\scalebox{0.3}{\includegraphics{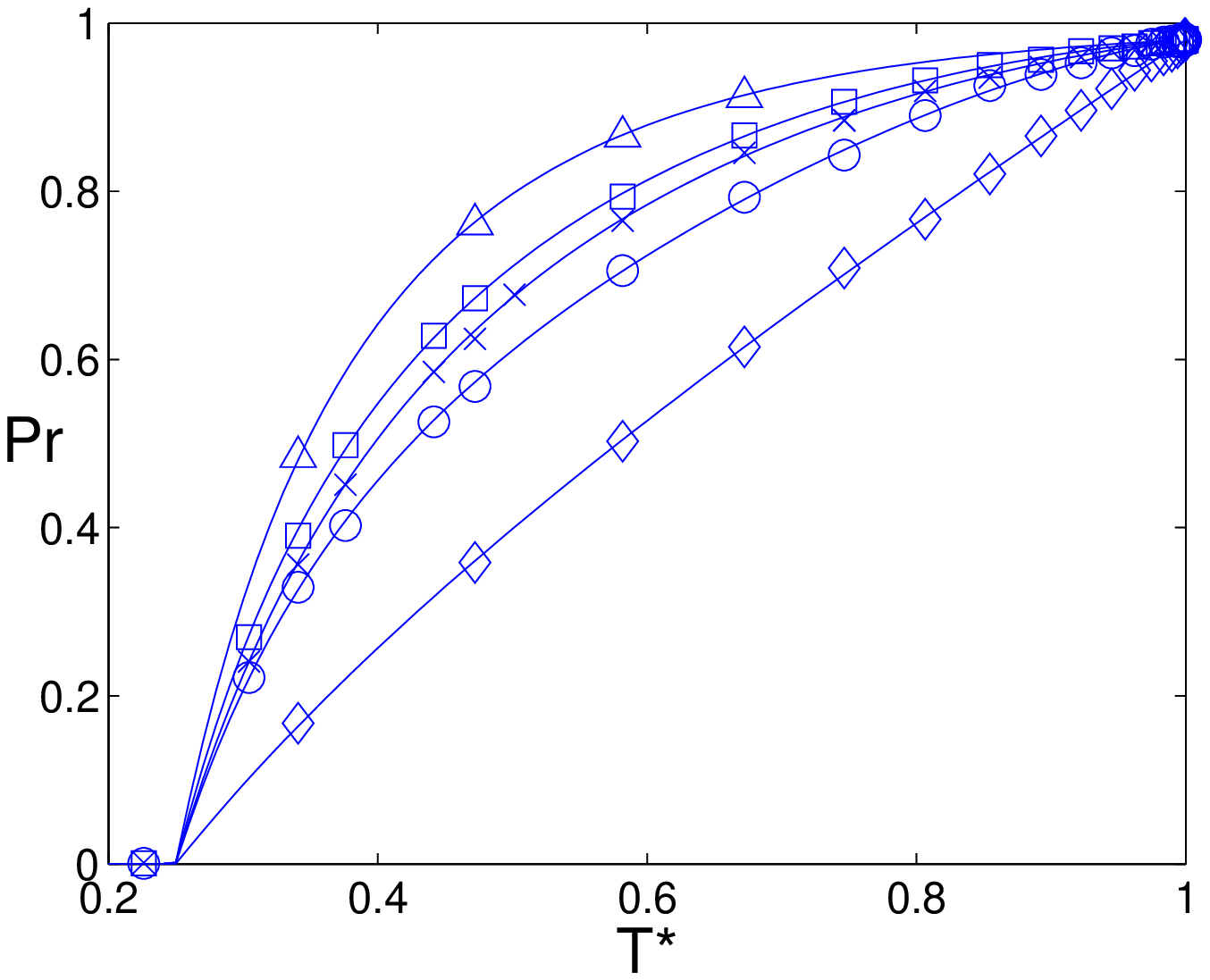}}\hfill
\scalebox{0.3}{\includegraphics{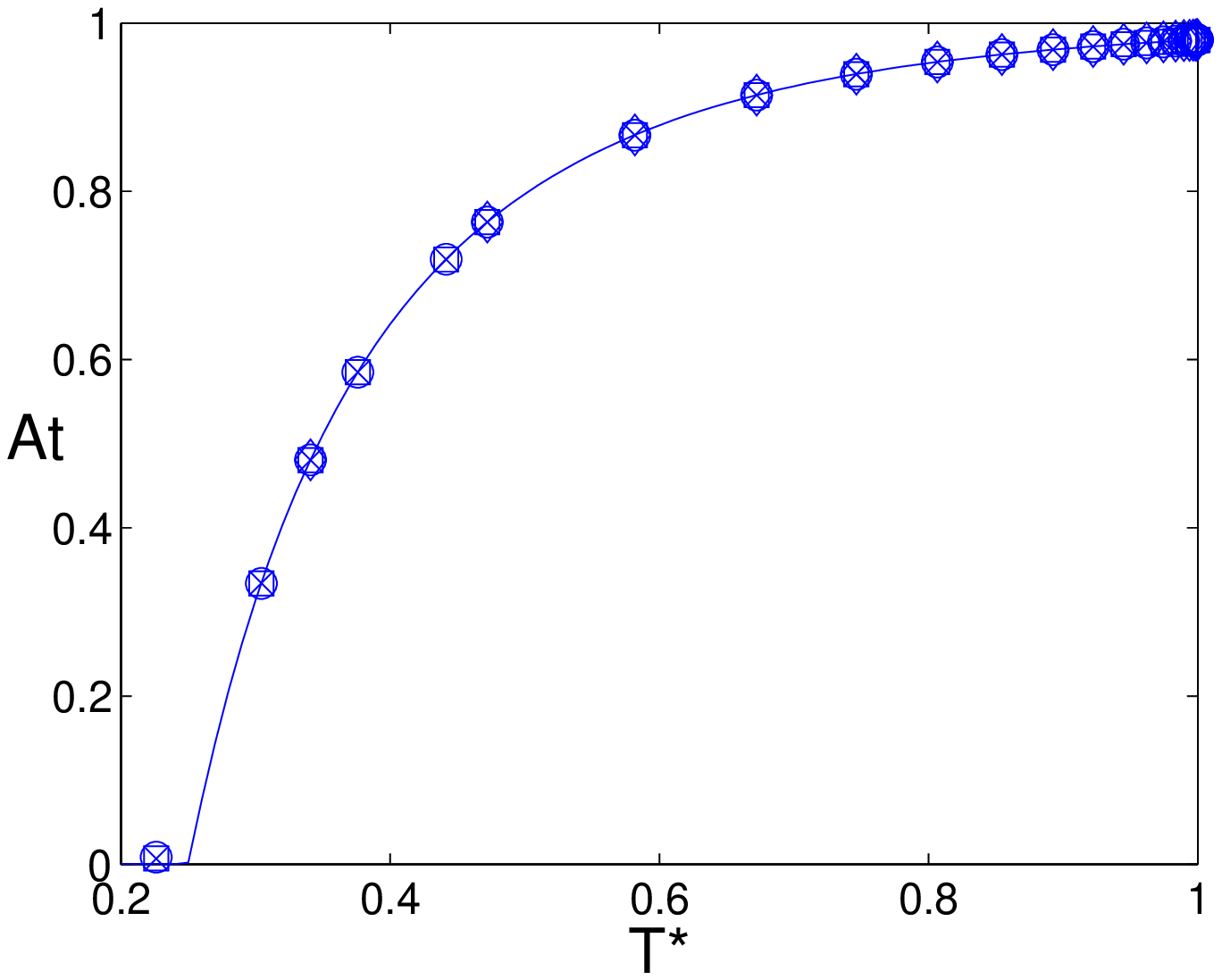}}
\caption{Comparison of theory (lines) with simulation (symbols).  For
  the different distributions of infectivity (with susceptibility
  constant), $\PE$ changes, but $\A$ does not.  We use constant
  recovery time $\tau = 5$ ($\triangle$),  $\tau=0$ or
  $\infty$ ($\lozenge$),  $\tau = 2$ or $8$ ($\square$),
  $\tau = 1$ or $10$ ($\circ$), and finally a constant
  recovery rate ($\times$).}
\label{fig:varyInf}
\end{figure}

For our first comparison, we consider the effect of varying infection
time.  We discretize time, and take different models of recovery time
given in the caption of figure~\ref{fig:varyInf}.  For each time step,
the probability of infecting a susceptible neighbor is $p$, and so an
infected individual with recovery time $\tau$ has $T_o=1-(1-p)^\tau$.
As a reference we consider the case where all infectious individuals
recover after exactly five time steps.  We vary $p$ in order to change
the average transmissiblity $T^*$.  The fraction of nodes with each recovery
time is chosen such that $\sum P(\tau)[1-(1-p)^\tau] = 1-(1-p)^5 =
T^*$.

The results of several examples are shown in figure~\ref{fig:varyInf}.
Each data point represents $10000$ simulations.  Away from the
epidemic threshold, there is a clear distinction between an epidemic
and a non-epidemic outbreak.  For definiteness, we define an epidemic to occur
if over $500$ nodes are infected.  Theory and simulations are in good
agreement.  The upper bound for epidemic probability is realized by
the case where all infections last exactly five time steps.  The lower bound
is realized by the case where some infections last forever and
infect all neighbors, while the rest recover before
infecting anyone.  Because susceptibility is homogeneous, $\A$ does not vary.


\begin{figure}
\scalebox{0.3}{\includegraphics{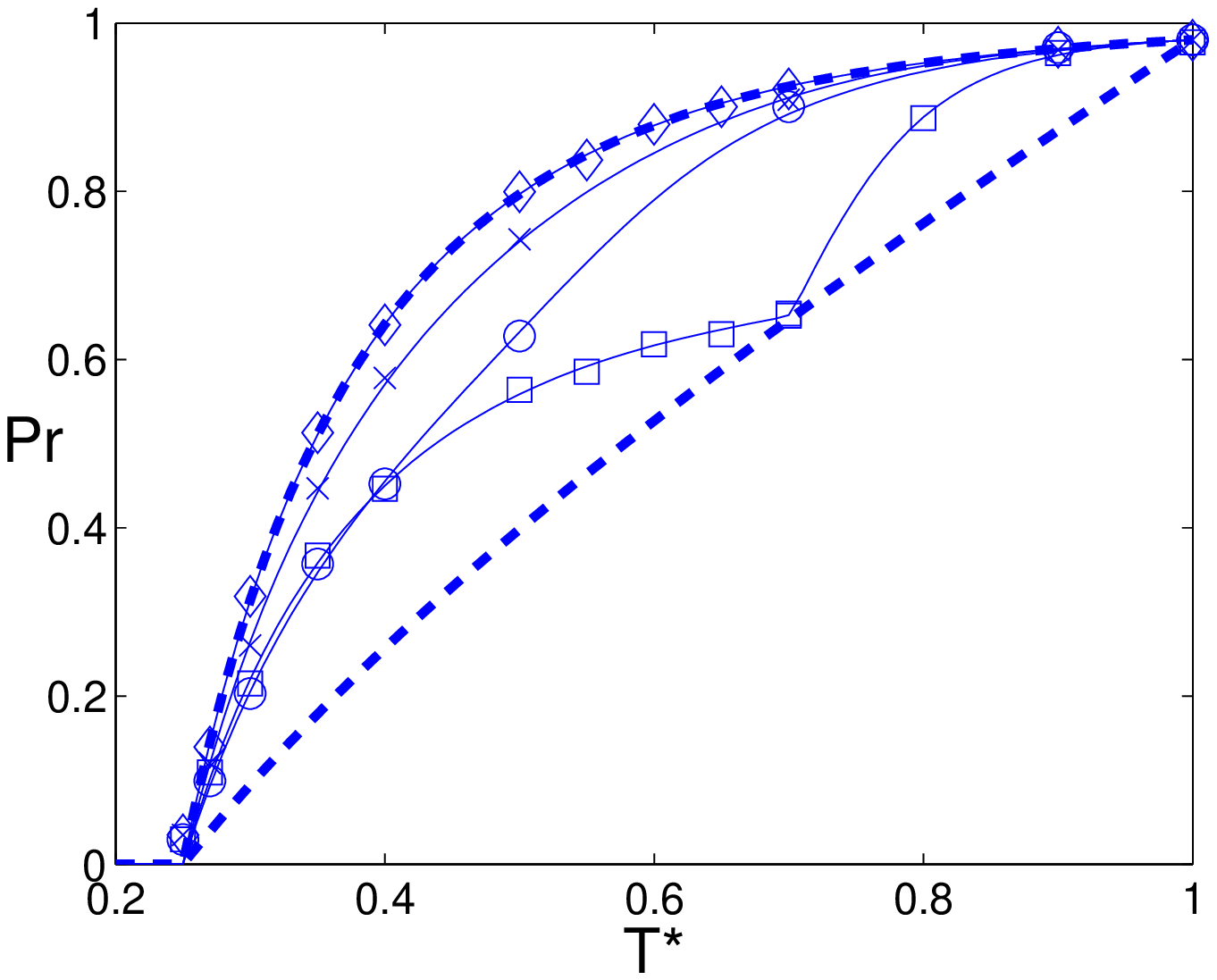}}\hfill
\scalebox{0.3}{\includegraphics{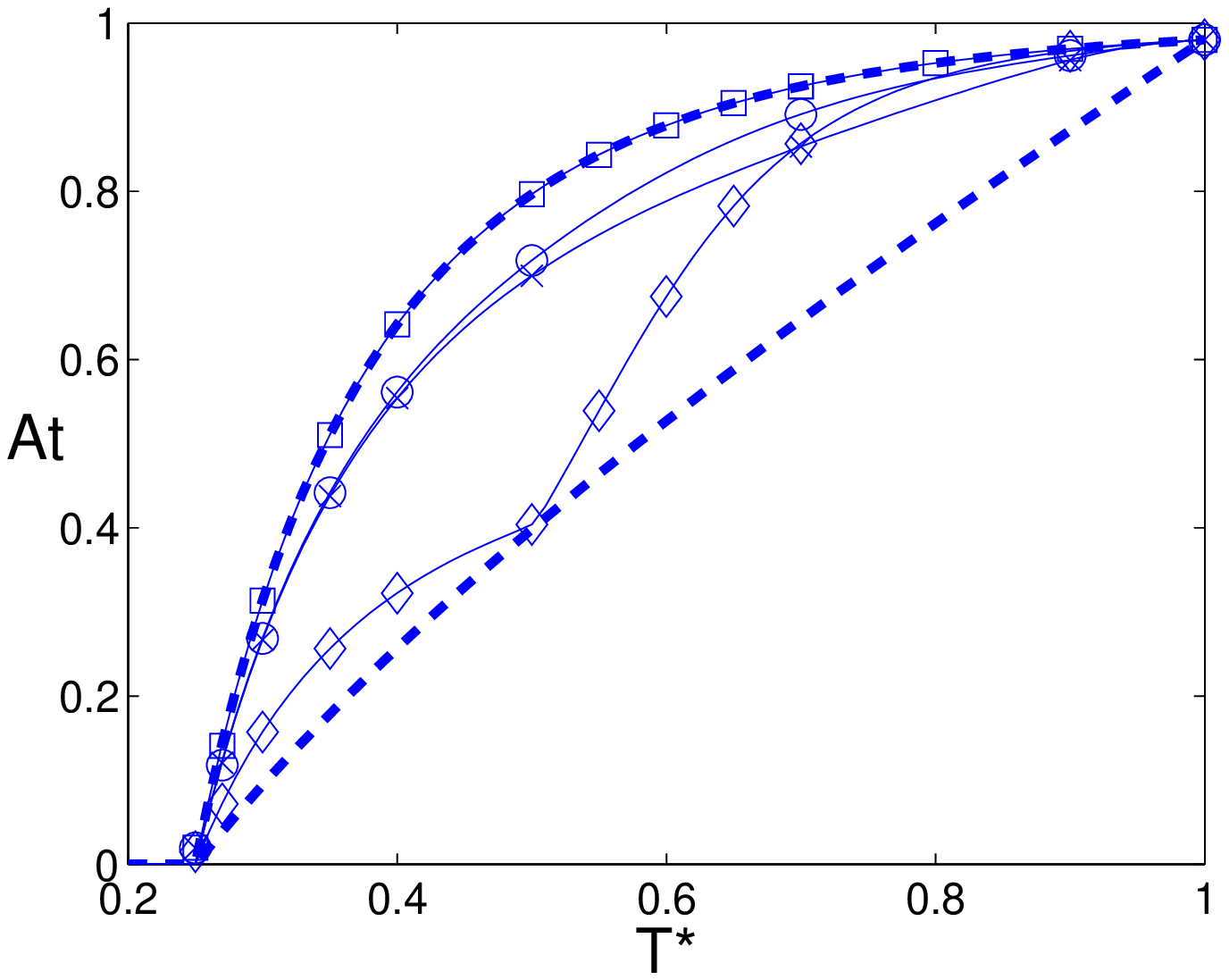}}
\caption{Comparison of theory (curves) with simulation (symbols) for
  $T_{uv} = 1-\exp(-\alpha \I_u \Sus_v)$.  The theoretical bounds are
  in dashed bold.  The distributions are $\lozenge$: $P(\I) =
  \delta(\I-1)$, \ $P(\Sus) = 0.5\delta(\Sus-0.001) +
  0.5\delta(\Sus-1)$; $\times$: $P(\I) =
  0.5\delta(\I-0.3)+0.5\delta(\I-1)$, \ $P(\Sus) =
  0.2\delta(\Sus-0.1)+0.8\delta(\Sus-1)$; $\circ$: $P(\I) =
  0.5\delta(\I-0.1)+0.5\delta(\I-1)$, \ $P(\Sus) =
  0.2\delta(\Sus-0.1)+0.8\delta(\Sus-1)$; $\square$: $P(\I) =
  0.3\delta(\I-0.001) + 0.7\delta(\I-1)$, \ $P(\Sus) =
  \delta(\Sus-1)$.}
\label{fig:varyBoth}
\end{figure}

For our second comparison we perform calculations for systems with
both $\I$ and $\Sus$ heterogeneous.  We use the same \erdosrenyi{}
network, but assume that $T_{uv} = 1-\exp[-\alpha \I_u \Sus_v]$ where
the distributions of $\I$ and $\Sus$ are fixed and $\alpha$ is varied
to tune $T^*$.  We find good agreement between
theory and simulations in
figure~\ref{fig:varyBoth}.

We have shown in this Letter that the effect of general heterogeneity
in infectivity and susceptibility on epidemic probability $\PE$ and
attack rate $\A$ can be accurately modeled using a generating function
approach.  We find that $\PE$ and $\A$ may differ substantially.  We
have further shown that heterogeneity in recovery time has a
significant effect on $\PE$ and cannot be ignored.

For fixed average transmissibility we have found upper and lower
bounds for both $\PE$ and $\A$.  Further we have found distributions
realizing these bounds.  For fixed average transmissibility,
increasing the variance of $P_o$ decreases $\PE$ and increasing
the variance of $P_i$ decreases $\A$.  

These results can be used to assist in designing control strategies.
For example, if choosing between a strategy that reduces infectivity
or susceptibility by half for all of the population or one that
reduces infectivity or susceptibility completely for half the
population, it is better to choose the latter.  As another example,
consider a strategy which attempts to locate and isolate infecteds
compared with a strategy which attempts to provide susceptible
individuals with protection.  Both may be affected by inability to
reach everyone.  The first strategy has a heterogeneous impact on
infectivity, while the second strategy has a heterogeneous impact on
susceptibility.  If the strategies have the same average impact on $T$
then the first reduces the probability of an epidemic more while the
second reduces its size more.  Which strategy is optimal depends on
whether the outbreak is small enough that an epidemic can be
prevented.

Our results can also be used to identify populations most at risk from
epidemics.  Populations with low genetic diversity are already known
to be at particularly high risk from an outbreak because the lack of
heterogeneity allows the transmissibility to be higher.  However, our
results show that even for a fixed average transmissibility, a
population with lower genetic variation will be more severely affected
by a disease.

For heterogeneous infectivity but homogeneous 
susceptibility, Newman~\cite{newman:spread} anticipated that $\A$
follows from the formulae derived under the assumption of homogeneous
$T$.  He did not address the effect on $\PE$.  We have shown that $\A$
is independent of heterogeneity in infectiousness, and so for this
special case the prediction is valid.  However, it fails if
susceptibility is also heterogeneous.

The theory developed here can be generalized in a number of ways.
Most simply, we can introduce edge weights to represent some details
of the contact between $u$ and $v$.  The same theory will hold, but
the calculation of $T_o$ and $T_i$ as in~\eqref{eqn:To} must
incorporate the edge weight distribution.  We can also introduce
correlations between the distributions of $\I$, $\Sus$, and $k$ in an
individual without significant theoretical difficulties, though the
conclusions may change.  It is more complicated to introduce
correlations of $\I$, $\Sus$, or $k$ between neighbors.

We have assumed that the network has few short cycles.  More realistic
social network models incorporate significant clustering.  However, at
high transmissibilities, if any neighbors are infected, an epidemic is
very likely.  $\PE$ is close to the probability that the initial
node infects any neighbors and loops may be ignored.  At low
transmissibilities loops are not traced out by the infection and again
may be neglected.  Loops affect our results only at intermediate
transmissibilities.  The generating function approach becomes
difficult because even early in an outbreak an infected node may have
multiple infected neighbors.

This work was carried out under the auspices of the National Nuclear
Security Administration of the U.S. Department of Energy at Los Alamos
National Laboratory under Contract No.\ DE-AC52-06NA2539.  The author
thanks Lindi Wahl, J. Mac Hyman, Anja C. Slim, Lu\'{i}s M. A.  Bettencourt,
Eduardo L\'{o}pez, Shweta Bansal, and Lauren A. Meyers for useful
discussions.

\bibliography{Joel_Miller_05Feb06}
\end{document}